\documentclass[12pt,a4paper]{article}

\usepackage{amsmath}
\usepackage{amssymb}
\usepackage[nosort]{cite}
\usepackage[hyperref,parsep]{collect}

\makeatletter
\let\old@startsection=\@startsection
\renewcommand{\@startsection}[6]{\old@startsection{#1}{#2}{#3}{#4}{#5}{#6\mathversion{bold}}}
\makeatother

\makeatletter
\let\old@makecaption=\@makecaption
\def\@makecaption{\small\old@makecaption}
\makeatother

\let\oldPhi=\Phi
\let\oldPsi=\Psi
\let\oldGamma=\Gamma
\let\oldDelta=\Delta
\let\oldSigma=\Sigma
\let\oldTheta=\Theta
\let\oldPi=\Pi
\renewcommand{\Phi}{\mathnormal{\oldPhi}}
\renewcommand{\Psi}{\mathnormal{\oldPsi}}
\renewcommand{\Gamma}{\mathnormal{\oldGamma}}
\renewcommand{\Sigma}{\mathnormal{\oldSigma}}
\renewcommand{\Delta}{\mathnormal{\oldDelta}}
\renewcommand{\Theta}{\mathnormal{\oldTheta}}
\renewcommand{\Pi}{\mathnormal{\oldPi}}

\newcommand{\superN}{\mathcal{N}}

\newcommand{\order}[1]{\mathcal{O}(#1)}

\newcommand{\Reals}{\mathbb{R}}

\newcommand{\cJ}{\mathcal{J}}

\ifx\genfrac\sdflkaj

\else

\fi
\newcommand{\sfrac}[2]{{\textstyle\frac{#1}{#2}}}
\newcommand{\half}{\sfrac{1}{2}}

\newcommand{\indup}[1]{_{\mathrm{#1}}}

\newcommand{\supup}[1]{^{\mathrm{#1}}}

\newcommand{\alg}[1]{\mathfrak{#1}}

\newcommand{\lrbrk}[1]{\left(#1\right)}
\newcommand{\bigbrk}[1]{\bigl(#1\bigr)}

\newcommand{\nl}[1][0pt]{\nonumber\\[#1]&\hspace{-4\arraycolsep}&\mathord{}}

\newcommand{\earel}[1]{\mathrel{}&\hspace{-2\arraycolsep}#1\hspace{-2\arraycolsep}&\mathrel{}}
\newcommand{\eq}{\earel{=}}

\def\[{\begin{equation}}
\def\]{\end{equation}}
\def\<{\begin{eqnarray}}
\def\>{\end{eqnarray}}

\makeatletter
\def\mr@ignsp#1 {\ifx\:#1\@empty\else #1\expandafter\mr@ignsp\fi}%
\newcommand{\multiref}[1]{\begingroup
\xdef\mr@no@sparg{\expandafter\mr@ignsp#1 \: }%
\def\mr@comma{}%
\@for\mr@refs:=\mr@no@sparg\do{\mr@comma\def\mr@comma{,}\ref{\mr@refs}}%
\endgroup}
\makeatother

\newcommand{\hypref}[2]{\ifx\href\asklfhas #2\else\href{#1}{#2}\fi}

\ifx\eqref\sdflkajd
\newcommand{\eqref}[1]{(\multiref{#1})}
\else
\renewcommand{\eqref}[1]{(\multiref{#1})}
\fi

\ifx\href\asklfhas\newcommand{\href}[2]{#2}\fi
\newcommand{\arxivno}[1]{\href{http://arxiv.org/abs/#1}{#1}}

\begin{document}
\thispagestyle{empty}
\renewcommand{\thefootnote}{\fnsymbol{footnote}}
\begin{flushright}\footnotesize
\texttt{\arxivno{hep-th/0509084}}\\
\texttt{PUTP-2175}\\
\texttt{NSF-KITP-05-73}
\end{flushright}
\vspace{1cm}

\begin{center}
{\Large\textbf{\mathversion{bold}%
On Quantum Corrections to\\Spinning Strings and Bethe Equations}\par}
\vspace{1cm}

\textsc{N.~Beisert$^{a,c}$ and A.A.~Tseytlin$^{b,c,}$%
\footnote{Also at Imperial College, London and Lebedev Institute, Moscow}}
\vspace{5mm}

\textit{$^a$ Joseph Henry Laboratories,
Princeton University\\
Princeton, NJ 08544, USA}\vspace{3mm}

\textit{$^b$ Department of Physics, The Ohio State University\\
Columbus, OH 43210, USA}\vspace{3mm}

\textit{\small $^{c}$ Kavli Institute for Theoretical Physics, University of California,\\
Santa Barbara, CA 93106, USA}\vspace{3mm}
\vspace{3mm}

\texttt{nbeisert@princeton.edu}\\
\texttt{tseytlin@mps.ohio-state.edu}\par\vspace{1.5cm}

\textbf{Abstract}\vspace{7mm}

\begin{minipage}{12.7cm}
Recently, it was demonstrated that one-loop energy shifts of 
spinning superstrings on $AdS_5\times S^5$ 
agree with certain Bethe equations for quantum strings
at small effective coupling.
However, the string result required artificial regularization by zeta-function.
Here we show that this matching is indeed correct up
to fourth order in effective coupling;
beyond, we find new contributions at odd powers.
We show that these are reproduced by quantum corrections within the Bethe ansatz.
They might also identify the ``three-loop discrepancy{}'' between string and gauge theory
as an order-of-limits effect.
\end{minipage}

\end{center}

\setcounter{footnote}{0}
\renewcommand{\thefootnote}{\arabic{footnote}}

\newpage
\setcounter{page}{1}
\setcounter{footnote}{0}


The investigation of semiclassical spinning superstrings on $AdS_5\times S^5$
\cite{Gubser:2002tv,Frolov:2002av,Russo:2002sr,Minahan:2002rc,Frolov:2003qc,Frolov:2003xy}%
\footnote{See \cite{Tseytlin:2003ii} for a review on 
semiclassical spinning strings.}
and their AdS/CFT duals, local operators of $\superN=4$ SYM 
in the thermodynamic limit \cite{Beisert:2003xu,Beisert:2003ea},%
\footnote{See \cite{Beisert:2004ry,Beisert:2004yq,Zarembo:2004hp} for reviews
on $\superN=4$ gauge theory and the thermodynamic limit.}
has lead to a number of important insights into both theories. 
Progress in this subject went hand in hand with the
discovery and development of integrable structures
in $\superN=4$ SYM \cite{Minahan:2002ve,Beisert:2003tq,Beisert:2003yb,Beisert:2003ys}
\ifx\nocollect\sdakjfhas\else\nocollect{Beisert:2003tq}\fi
\ifx\nocollect\sdakjfhas\else\nocollect{Minahan:2002ve}\fi
\ifx\nocollect\sdakjfhas\else\nocollect{Beisert:2003yb}\fi
and string theory on $AdS_5\times S^5$ \cite{Mandal:2002fs,Bena:2003wd}.%
\footnote{See \cite{Beisert:2004ry,Beisert:2004yq,Zarembo:2004hp} for reviews
on integrability of gauge theory and strings.}
The computations of the spinning string correspondence 
required powerful methods which integrability could provide.
Conversely, spinning strings were an ideal testing ground
for these methods. 

The main tool for obtaining the spectrum of
integrable models is the Bethe ansatz. For gauge theory it was 
developed in \cite{Minahan:2002ve,Beisert:2003yb,Serban:2004jf,Beisert:2004hm,Staudacher:2004tk,Beisert:2005fw}.
The string counterpart is a set of integral equations 
for classical strings \cite{Kazakov:2004qf,Kazakov:2004nh,Beisert:2004ag,Beisert:2005bm}
and a proposal for the promotion to Bethe equations for quantum strings 
was made in \cite{Arutyunov:2004vx,Staudacher:2004tk,Beisert:2005fw}.
The comparison of the classical spectra of both models has
shown general agreement at the leading two orders 
\cite{Beisert:2003xu,Beisert:2003ea,Kazakov:2004qf,Schafer-Nameki:2004ik,Beisert:2005di,Beisert:2005fw}, 
but also lead to the discovery of a disagreement at third order \cite{Serban:2004jf},%
\footnote{See also \cite{Callan:2003xr} for a
closely related effect in the near plane wave/BMN correspondence.}
the so-called ``three-loop discrepancy{}''.%
\footnote{See \cite{Tseytlin:2004xa,Plefka:2005bk} for reviews
on the comparison between semiclassical spinning strings and
gauge theory.}
Note that this mismatch is not necessarily in conflict with the AdS/CFT correspondence
\cite{Maldacena:1998re,Gubser:1998bc,Witten:1998qj}
though, because order-of-limits effects 
may spoil the (naive) comparison \cite{Serban:2004jf,Beisert:2004hm}.
\smallskip

Recently, the precision tests of the quantum string Bethe equations 
were performed by comparing their prediction to one-loop effects 
in quantum string theory.
String energies $E(\lambda,J)$ admit an expansion 
for large string tension $\sqrt{\lambda}$ (or large 't~Hooft coupling $\lambda$)
%
\[\label{eq:EnergyExpand}
E(\lambda,J)=\sqrt{\lambda}\,\mathcal{E}(\cJ)+\delta E(\cJ)+\order{1/\sqrt{\lambda}},
\qquad \cJ={J}/{\sqrt{\lambda}}\,.
\]
Here, $\mathcal{E}$ is the classical string energy and $\delta E$ is 
the one-loop energy shift.
The effective string tension $\tilde \lambda^{1/2}=1/\cJ$
(alias the effective spin $\cJ$) can take any fixed value. 
The comparison of $\delta E$ was performed in an expansion in powers of
the effective coupling $1/\cJ$.
Agreement at $\order{1/\cJ^2}$ for the simplest class of spinning string solutions 
was found in \cite{Beisert:2005mq,Hernandez:2005nf}. 
This was later generalized to the full $\alg{su}(2)$ sector \cite{Beisert:2005bv}.
Going to higher orders in $1/\cJ$, however, is problematic
due to the appearance of divergent sums \cite{Park:2005ji}.
When these sums are regularized by the first regulator that might come to mind,
namely by zeta-function, the result does indeed agree with the 
Bethe ansatz at $\order{1/\cJ^6}$ \cite{Schafer-Nameki:2005tn}.
This is a very good sign of the validity of the Bethe ansatz, 
given that the computation and the result are rather complex.
Merely the need to regularize 
within this conformal two-dimensional model
appears artificial; 
the unexpanded sums do indeed converge \cite{Frolov:2004bh}. 

\medskip

In this note we investigate the divergent sums carefully 
and find that one can make sense of them. 
This allows us to compute the coefficients of the
expansion of the one-loop energy shift $\delta E$. 
We find that zeta-function regularization actually produces the
correct coefficients of $1/\cJ^4$ and $1/\cJ^6$. 
However, we find additional contributions at \emph{odd} powers of $1/\cJ$ 
starting at $\order{1/\cJ^5}=\order{\lambda^{5/2}/J^5}$.%
\footnote{Similar observations are made in \cite{Schafer-Nameki:2005is}.
There, systematic analytic methods of handling sums and of 
computing corrections were developed on
several examples. Their methods may be more suitable 
to understand potential exponential corrections
beyond the perturbation series.}
This may appear disastrous for the quantum string Bethe ansatz,
which does not produce such terms, and for the comparison to gauge
theory, due to the unexpected fractional powers of $\lambda$.
Nevertheless, quite the contrary is true: On the one hand, 
we will demonstrate that
these contributions allow us to determine quantum corrections 
to the Bethe equations themselves.
That this is possible at all is non-trivial and therefore makes
us more confident of the Bethe ansatz for quantum strings. 
On the other hand, they can be interpreted as large-$\lambda$
effects which might repair the disagreement between string and gauge theory
when interpolated down to small $\lambda$. 
Here we even see some quantitative confirmation of this idea.

\bigskip

Let us now reinvestigate the one-loop energy shift 
of a circular spinning string on $AdS_3\times S^1$
in string theory. 
The classical solution was  found in \cite{Arutyunov:2003za}
and quantum corrections to the energy were computed in \cite{Park:2005ji}.
We will use the notation of \cite{Beisert:2005mq,Schafer-Nameki:2005tn},
i.e.~$k$ is the mode number, $m$ is the winding number for $S^1$ and
$n$ is the mode number of the fluctuation.
The spin $S$ on $AdS_3$ and the spin $J$ on $S^1$ are related by
$S k+J m=0$.
The energy shift is given by the generic formula
\[\label{eq:SumOfModes}
\delta E=\sum_{n=-\infty}^{\infty} e(n),
\]
where $e(n)$ is the sum of 
contributions of bosonic and fermionic fluctuations with given mode number $n$.
The expression $e(n)$ can be found in \cite{Park:2005ji,Beisert:2005mq,Schafer-Nameki:2005tn},
we recall it in \eqref{eq:eofn} in the appendix.
\smallskip

We first expand for large $\cJ$ at fixed $n$
and denote the result by $e\supup{sum}(n)$.
It then turns out that 
starting from $\order{1/\cJ^4}$ 
the sum of $e\supup{sum}(n)$ diverges 
due to contributions with positive powers of $n$.%
\footnote{We sum order by order in $1/\cJ$. 
Technically, the divergencies are caused by an order-of-limits effect.} 
Let us therefore split the result into a regular part 
$e\supup{sum}\indup{reg}(n)$
with contributions of $\order{1/n^2}$ 
and a singular part 
$e\supup{sum}\indup{sing}(n)$
polynomial in $n$.
The expressions are lengthy and we present them 
in eqs.  \eqref{eq:SumReg},\eqref{eq:SumSing} in the appendix.
Clearly, the sum of the regular part converges while
the sum of the singular part, an even polynomial, 
gives identically zero when regularized by zeta-function. 
For small values of $n$ our answer appears fine, 
but the large-$n$ behavior is incompatible with the expansion.
This problem is not unexpected as we have assumed $n$ to be fixed 
while taking $\cJ$ large. 
This very assumption conflicts with the nature of the sum 
which goes over \emph{all} modes $n$.
\smallskip

Let us now attempt to improve the approximation for large values of $n$. 
For this we set $n=\cJ x$ and expand for large $\cJ$.
The resulting expression is given in \eqref{eq:IntFull} in the appendix.
In this case, the energy shift should be approximated by 
the integral of $\cJ dx\,e\supup{int}(x)$.
Once again, we find a problem:
The integrand diverges at $x=0$, as was already noticed
for a similar solution on $\Reals\times S^3$ in \cite{Frolov:2004bh},
and the integral cannot be performed. 
To see more clearly what happens, we separate the integrand
into a regular part $e\supup{int}\indup{reg}(x)$
which is smooth at $x=0$ 
and a singular part $e\supup{int}\indup{sing}(x)$ with strictly positive powers of $1/x$.
The singular part is given in eq.\eqref{eq:IntSing} in the appendix.
Despite the singularities at $x=0$, 
let us note that $e\supup{int}(x)$ 
has the correct asymptotics at large $n$, c.f.~\eqref{eq:IntAsymp};
its expansion agrees quantitatively with the asymptotics of $e(\cJ x)$.
Apparently, here $e\supup{int}(x)$ approximates
$e(n)$ well at large values of $n=\cJ x$, but not at small ones.
\smallskip

The divergencies at large $n$ in the first approach 
are traded in for divergencies at small $n$ in the second one.
We might therefore try to combine the two approaches, 
use $e\supup{sum}(n)$ for small $n$ and 
$e\supup{int}(x)$ for large $n$.
As we will see, this can be done. Moreover,
we do not even need a cut-off to separate between the two regimes. 
Instead we make use of the following observation:
The singular part in one regime seems to equal the
regular part in the other regime:
$e\supup{int}\indup{sing}(x)=e\supup{sum}\indup{reg}(\cJ x)$
and $e\supup{sum}\indup{sing}(n)=e\supup{int}\indup{reg}(n/\cJ)$.
This property can be confirmed by expanding 
the regular part after interchanging $n$ and $\cJ x$.%
\footnote{In physicist's terms:
\emph{resumming} one singular part yields the other regular part.}
We thus find%
\footnote{Note that $e\supup{int}\indup{sing}$ and $e\supup{sum}\indup{reg}$
have positive powers of $x,n$
while $e\supup{sum}\indup{sing}$ and $e\supup{int}\indup{reg}$ 
have strictly negative ones.
Consequently, we might understand this split as
a Laurent expansion in $n$ and a subsequent 
separation into positive and strictly negative powers.}
\[
e(n)=
e\supup{sum}\indup{reg}(n)
+e\supup{int}\indup{reg}(n/\cJ)
=
e\supup{sum}\indup{sing}(n)
+e\supup{int}\indup{sing}(n/\cJ).
\]
Therefore, there is no need to consider the singular parts at all;
to obtain the energy shift it suffices to consider the regular parts%
\footnote{The integral is merely an approximation to the sum.
We however did not find any corrections 
polynomial in $1/\cJ$ by improving the 
integrand using the Euler-Maclaurin formula as in \cite{Frolov:2003tu}.}
\[
\delta E=\sum_{n=-\infty}^{\infty}e(n)=
\sum_{n=-\infty}^{\infty}e\supup{sum}\indup{reg}(n)
+\int_{-\infty}^{\infty} \cJ dx\, e\supup{int}\indup{reg}(x).
\]
The sum of $e\supup{sum}\indup{reg}(n)$ is known,
it is the zeta-function regularized sum in \cite{Schafer-Nameki:2005tn}.
The integral however yields a non-trivial contribution
\[\label{eq:NewTerms}
\int_{-\infty}^{\infty} \cJ dx\, e\supup{int}\indup{reg}(x)=
-\frac{(k-m)^3 m^3}{3\cJ^5}
+\order{1/\cJ^7}.
\]
It is somewhat surprising to see that
the integrand $e\supup{int}\indup{reg}(x)$,
c.f.~\eqref{eq:IntFull},\eqref{eq:IntSing}, starts
at $\order{1/\cJ^4}$, but its integral vanishes at this order. 
Nevertheless, this is merely an exception, 
the integral does not vanish at higher orders. 
While all the contributions from 
$e\supup{sum}\indup{reg}(n)$ are at even powers of $1/\cJ$,
the new contributions are at odd powers. 
Put differently, the first new term is at order $\lambda^{5/2}/J^5$.

\medskip

The new term in \eqref{eq:NewTerms} contradicts
the naive expectation that the expansion goes in integer
powers of $\lambda$ and $1/J$ \cite{Frolov:2003tu}
and thus can be directly compared
to perturbative gauge theory.
It also contradicts the simplest version of 
the Bethe ansatz for quantum strings \cite{Arutyunov:2004vx,Staudacher:2004tk,Beisert:2005fw} 
which does not produce  such terms \cite{Schafer-Nameki:2005tn}.
Nevertheless, the appearance of such terms leads to 
a natural proposal of how to establish the agreement between 
the gauge and string theory results. 

First of all, the one-to-one comparison 
of perturbative string theory to perturbative gauge theory
is suggestive but seemingly plagued by order-of-limits effects. 
On top of the well-known disagreement of coefficients, 
the ``three-loop discrepancies{}'' \cite{Callan:2003xr,Serban:2004jf}, 
here we find that also the \emph{structure} 
of the expansion is different in both limits.
This is not in conflict with AdS/CFT;
it merely invalidates attempts to compare perturbatively.

Let us assume the AdS/CFT correspondence holds.
Then the exact energy $E$ should be some interpolating function between 
the perturbative string theory expression at large $\lambda$ 
and the perturbative gauge theory at small $\lambda$.
Now we note that the new term at $\order{1/\cJ^5}$ 
is accompanied by an old term at the same order in $1/\cJ$
coming from the expansion of the classical string energy 
at $\order{\sqrt{\lambda}/\cJ^5}$.
The former should be considered as a quantum correction to the latter.
We might combine these two terms with higher-loop
corrections into some function $f_5(\lambda)\sqrt{\lambda}/\cJ^5$
of the coupling. 
At large $\lambda$ the function $f_5(\lambda)$ admits an expansion in powers
of $1/\sqrt{\lambda}$ starting at $\order{1}$;
here we merely see the first two terms.
At small $\lambda$ we expect $f_5(\lambda)$ to have a regular
expansion in $\lambda$.
In between, it should interpolate between $f_5(\infty)$ and $f_5(0)$.
Similar effects have been observed in 
the related context of plane wave string field theory in \cite{Klebanov:2002mp}.

In fact, it is precisely this term, 
$\order{\sqrt{\lambda}/\cJ^5}$ in  string theory  and 
$\order{\lambda^3/J^5}$ in gauge theory,
at which the three-loop discrepancy starts \cite{Serban:2004jf}.
Put differently, we find $f'_p(\infty)\neq 0$ precisely for that value 
of $p$ where $f_p(\infty)\neq f_p(0)$.
This might be a sign that the mismatch will be resolved 
by an interpolation between strong and weak coupling.%
\footnote{Similar qualitative statements appeared in 
\cite{Serban:2004jf,Beisert:2004hm,Arutyunov:2004vx,Tseytlin:2004xa,Mann:2005ab},
see also \cite{Klebanov:2002mp}.}
Below, we will present some quantitative evidence
for this qualitative statement.

\medskip

How can the new contribution be interpreted in the
quantum string Bethe ansatz \cite{Arutyunov:2004vx,Staudacher:2004tk,Beisert:2005fw}? 
According to the sophisticated analysis in 
\cite{Schafer-Nameki:2005tn}, 
the expansion of $\delta E$ is in even powers
of $1/\cJ$, at least up to $\order{1/\cJ^6}$.
Here we go back to the original proposal 
of the string Bethe equations in \cite{Arutyunov:2004vx}. 
Arutyunov, Frolov and Staudacher's proposal 
was to modify the gauge Bethe equations by an additional
phase shift for the interchange of two excitations%
\footnote{In the proposal of \cite{Arutyunov:2004vx},
the interpolating functions could also depend on spin $J$ or length $L$.
This might seem somewhat unnatural from a Bethe ansatz/spin chain/S-matrix point of view. 
Furthermore, it is not clear how to define $L$ in string theory.
Indeed, we will not need dependence on $L$.}
\[\label{eq:Dressing}
\theta(p_k,p_j)=2\sum_{r=2}^\infty c_r(\lambda)\,(\lambda/16\pi^2)^{r} \bigbrk{q_r(p_k)\, q_{r+1}(p_j)-q_{r+1}(p_k)\, q_r(p_j)}.
\]
This dressing phase $\theta$ depends on the momenta $p$ of the excitations
through their conserved charges $q_r$.
The undetermined functions $c_r(\lambda)$ should approach $1$ 
at $\lambda\to\infty$ to obtain the correct classical limit.
If they interpolate to $0$ at $\lambda=0$, the Bethe equations might
even agree with the correct gauge result.
Apart from these two limits, we know no further constraints for the $c_r$ yet.
In \cite{Schafer-Nameki:2005tn} it was assumed that the functions
$c_r(\lambda)=1$ are exact, 
i.e.~they do not receive string quantum corrections;
that led  to an expansion of the string energy in even powers of $1/\cJ$.
\smallskip

Let us now see whether we can re-establish agreement with 
one-loop string theory by correcting $c_2=1+\epsilon$.
We thus add an overall phase to the Bethe equations%
\footnote{We generalize the form of the corrections
to include two uncorrelated charges $q_r$ and $q_s$.
This appears to be the most general form for 
Bethe equations for certain types of spin chains
\cite{Klose:2005aa,Beisert:2005aa}.
We thank T.~Klose and M.~Staudacher for discussions. 
}
\[
2\epsilon (\lambda/16\pi^2)^{(r+s-1)/2} \bigbrk{q_r(p_k)\, q_s(p_j)-q_s(p_k)\, q_r(p_j)}.
\]
We solve the Bethe equations for the $\alg{sl}(2)$ sector 
in the thermodynamic limit with all mode numbers coinciding. 
This is the one-cut solution studied in \cite{Kazakov:2004nh,Beisert:2005mq,Hernandez:2005nf,Schafer-Nameki:2005tn}
corresponding to the above circular spinning string.
The equations can be solved by the standard trick of turning
them into a quadratic equation for a resolvent. 
We then find that the classical energy shifts by%
\footnote{This is the result for the $\alg{su}(2)$ Bethe
equations. The result for $\alg{sl}(2)$ is similar.}
\[\label{eq:BetheShift}
\delta \mathcal{E}=4\epsilon \,
  \frac{\mathcal{Q}_{r+1}\mathcal{Q}_{s}-\mathcal{Q}_{r}\mathcal{Q}_{s+1}}
  {(4\pi)^{r+s+1}\mathcal{E}}
+\order{\epsilon^2}.
\]
Here $\mathcal{Q}_r$ are the conserved charges of the solution
as defined in \cite{Beisert:2004hm}, 
here they are normalized to scale as $\order{1/\cJ^{r-1}}$,
c.f.~\cite{Arutyunov:2003rg}.
We find for the energy shift with $r=2$ and $s=3$
\[\label{eq:ShiftSL2}
\delta \mathcal{E}=\epsilon\,\frac{(k-m)^3 m^3}{16\cJ^5}
+\order{1/\cJ^7}.
\]
Remarkably,  this is in structural agreement with \eqref{eq:NewTerms}.
When we set in \eqref{eq:Dressing}
\[\label{eq:C2}
c_2(\lambda)=1-\frac{16}{3\sqrt{\lambda}}+\order{1/\lambda}
\]
the Bethe equations reproduce the correct string result for our class
of circular solutions parametrized by $k,m$.
\medskip


In fact, one can easily convince oneself that the leading 
discrepancy between classical string energies $E\indup{s}$ and 
gauge theory energies in the thermodynamic limit $E\indup{g}$
is obtained from \eqref{eq:ShiftSL2} for $\epsilon=1$. 
The general prediction for the $\order{\lambda^{5/2}}$ contribution 
of an arbitrary solution is thus $-\frac{16}{3}(E\indup{s}-E\indup{g})/\sqrt{\lambda}$.
So our finding is completely consistent with the idea that
$c_2$ interpolates between $1$ at strong coupling
and $0$ at weak coupling. 
This suggests a natural resolution of the 
apparent disagreement between the string and 
gauge theory results at order $\lambda^3$
from a string perspective.

Conversely, each effect should have a counterpart 
on the other side of the duality. 
How can the disagreement be reduced from a gauge theory point of view?
This depends crucially on how the functions $c_r(\lambda)$ approach zero
near $\lambda=0$. 
For an exponential decline, such as $c_2(\lambda)=\exp(-\frac{16}{3}/\sqrt{\lambda})$, 
we would see no effects in perturbative gauge theory at all. 
Another possibility is that $c_r(\lambda)\sim \lambda^L$, where $L$ is the length
of the state.%
\footnote{
This would, however, be in contradiction with the philosophy
of a length-independent S-matrix.}
This behavior might be associated to ``wrapping effects{}''
\cite{Beisert:2004hm}, special types of corrections which start 
when the range of the Hamiltonian exceeds the length of the state.%
\footnote{Alternatively, the \emph{asymptotic} Bethe ansatz 
might break down at this order and needs to be replaced
by something structurally different from a Bethe ansatz,
see e.g.~\cite{Inozemtsev:2002vb}.}
If, however, $c_r(\lambda)\sim \lambda^a$ with some fixed $a$, 
then the scaling behavior in the thermodynamic limit 
(as well as BMN-scaling \cite{Berenstein:2002jq}) would break down. 
Proper scaling was a central assumption in
the construction of higher-loop gauge theory results
(see \cite{Beisert:2004ry} for a review), 
but has only been confirmed rigorously up to $\order{\lambda^3}$
\cite{Beisert:2003ys,Eden:2004ua}.%
\footnote{The dispersion relation would still 
preserve scaling as well as most parts of the S-matrix. 
Only a global phase would violate proper scaling.
A first guess $c_2(\lambda)\sim\lambda$ would imply  
scaling violations in gauge theory at four loops,
just slightly beyond our current horizon.
Intriguingly, such scaling violations have been
observed in the plane wave matrix model \cite{Fischbacher:2004iu}.
This latter fact does not necessarily have implications for $\superN=4$ SYM.}
\bigskip

Of course, the interpolating functions of the string Bethe ansatz, 
e.g.~\eqref{eq:C2}, must be universal
and hold for all other solutions in any sector as well
\cite{Staudacher:2004tk,Beisert:2005fw}.
We can thus predict the contributions at odd powers of $1/\cJ$
from the Bethe equations.
To see this, let us repeat the above analysis in a different sector,
for a circular string on $\Reals\times S^3$ \cite{Frolov:2003qc}.
This corresponds to the $\alg{su}(2)$ single-cut 
solution of \cite{Kazakov:2004qf,Minahan:2004ds}.%
\footnote{This solution is unstable due to 
tachyonic modes at small $n<2m$ (IR).
Here we consider corrections which are associated to 
large mode numbers (UV) and thus unaffected by the instability.}
We restrict to the ``half-filling{}'' point ($J_1=J_2=J/2$),
where most expressions simplify. 
For the corrections at odd powers of $1/\cJ$ we appear to find, 
using the expressions in \cite{Frolov:2004bh}%
\footnote{This result is independent of way periodicity is handled for fermions,
c.f.~\cite{Frolov:2003qc,Frolov:2004bh} vs.~\cite{Arutyunov:2003za,Beisert:2005mq}.}
\footnote{This result can also be obtained from string theory 
with an infinite world sheet confirming that 
the origin of the contribution is a local quantum effect
rather than a finite-size effect.}
\footnote{Comparing \cite{Arutyunov:2003za} and \cite{Minahan:2004ds}
we expect $-m^3(k-m)^3/3\cJ^5+\order{1/\cJ^7}$ for the generic case. Here $k=2m$.}
\[
\int_{-\infty}^{\infty} \cJ dx\, e\supup{int}\indup{reg}(x)=
\frac{m^2}{\sqrt{\cJ^2+m^2}}
+\frac{2\cJ^2}{\sqrt{\cJ^2+m^2}}\log\frac{\cJ^2}{\cJ^2+m^2}
-\frac{\cJ^2-m^2}{2\sqrt{\cJ^2+m^2}}\log \frac{\cJ^2-m^2}{\cJ^2+m^2}\,.
\]
This agrees with the Bethe equations when $c_2$ is as in \eqref{eq:C2}.%
\footnote{A preliminary analysis using \eqref{eq:BetheShift}
yields the leading corrections for the higher $c_r(\lambda)$. 
The coefficients for $c_2\ldots c_6$ seem to be:
$-16/3$, $-16/3$, $-184/15$, $-182/15$, $-3268/175$, \ldots without an apparent pattern.}
We have also performed a numerical comparison between the 
exact sum and our expansion of it. We set $m=1$ and 
sum up to $|n|=5000$ for $\cJ$ between $3$ and $10$.
The coefficients of the $1/\cJ$ expansion 
are evaluated numerically up to $\order{1/\cJ^9}$. 
The results of both approaches agreed up to about $10^{-7}$.
If,  however, we  eliminate the odd powers in $1/\cJ$ from the expansion, 
the matching is reduced to about $10^{-4}$. 
This is a clear verification of the presence  of the 
odd powers of $1/\cJ$ in the expansion. 
\medskip

There are other cases for which one might compute these odd contributions.
For instance, there are further one-cut solutions which should be 
easily accessible, such as solutions on $\Reals\times S^5$
\cite{Minahan:2002rc,Frolov:2003qc,Engquist:2003rn,Freyhult:2005fn}.
These are interesting because they add ``flavor{}'' to the Bethe equations.
One could also try to generalize to two-cut solutions, 
such as the folded string \cite{Beisert:2003xu,Frolov:2003xy}, 
but these are more involved due to their elliptic nature.
An expansion around an algebraic solution 
along the lines of \cite{Kristjansen:2004ei,Kristjansen:2004za}
might simplify the analysis.
\smallskip

The universality of the Bethe ansatz also predicts 
the existence of these types of corrections in 
the near plane wave limit of $AdS_5\times S^5$.
There, the first terms of fractional order in $\lambda$ would occur
at $\order{\lambda^{5/2}/J^7}$ representing a $1/J^2$ effect. 
At second order in $1/J$ a sum over intermediate channels appears
and this may become divergent when first expanding in 
$\lambda'=\lambda/J^2$.%
\footnote{A similar problem has been observed in 
the context of plane wave string field theory in \cite{Klebanov:2002mp}
when the expansion for small $\lambda'=\lambda/J^2$ 
is done prior to summing.}
Partial results were obtained in \cite{Swanson:2004mk}. 
Once the exact expressions for finite $\lambda'$ are known, 
the regularization of the sum might proceed in a 
similar way as above and the result should be compared to the
Bethe ansatz.
For instance, in the $\alg{su}(2)$ sector,
the leading difference between gauge and  
string theory in the near plane wave limit 
is given by the general formula derived from 
the results in \cite{Arutyunov:2004vx}%
\[
E\indup{s}-E\indup{g}=-\frac{\lambda^3}{16J^7}
\sum_{k,j=1}^M
n_k^2n_j^2(n_k-n_j)^2+\order{\lambda^4/J^9}.
\]
Here, $M$ is the number of excitations and $n_k$ are their mode numbers
(which are allowed to coincide).
The $\order{\lambda^{5/2}/J^7}$ contribution is predicted to be
${-\frac{16}{3}(E\indup{s}-E\indup{g})/\sqrt{\lambda}}$.%
\medskip

One might also wonder how to obtain the 
odd powers in $1/\cJ$ in the fast string expansion 
of \cite{Kruczenski:2003gt,Kruczenski:2004kw,Mikhailov:2004xw}.%
\footnote{See \cite{Tseytlin:2004xa} for a review
of the fast string expansion.}
Here, one expands in $1/\cJ$ at the level of the classical action.
Therefore,  one can possibly obtain only the summands $e\supup{sum}(n)$ 
expanded at finite mode number $n$. 
As we have demonstrated, the integrand $e\supup{int}(x)$ may be
recovered from $e\supup{sum}(n)$. However, this requires resumming
of all orders and thus the odd powers in $1/\cJ$ are non-perturbative
contributions in this effective field theory.
\medskip

There are many aspects which deserve further investigation.
For instance, it would be important to understand how to disentangle 
finite-size ($1/J$) and finite-tension ($1/\sqrt{\lambda}$) effects:
We have interpreted the odd powers in $1/\cJ$ 
as quantum corrections to classical contributions. 
They correspond to $1/\sqrt{\lambda}$ corrections to the Bethe equations.
Also, when extrapolating to perturbative gauge theory, 
these terms should go away. 
On the other hand, the corrections at even powers in $1/\cJ$ 
remain and can be compared to gauge theory. 
There, they correspond to finite-size ($1/J$) corrections
to the thermodynamic limit.
If we knew how to disentangle them, 
we could focus only on finite-tension effects 
and find higher loop corrections to the Bethe equations.%
\footnote{Related issues and ideas have been discussed in \cite{Mann:2005ab}
which might be useful in this respect.}
\smallskip

Another direction to proceed would be to generalize 
the findings of \cite{Beisert:2005bv} to finite $1/\cJ$. 
At $\order{1/\cJ^2}$ it was shown in generality that
the one-loop energy shift equals a regularized sum over fluctuation energies. 
As above, the regularization should be equivalent to
adding quantum corrections to the Bethe equations. 
Now, the fluctuation energies and the energy shift can both
be computed from the Bethe equations. 
By comparing the two, one should thus be able to derive 
the complete one-loop quantum corrections
as a consistency requirement of the Bethe ansatz framework
with quantum mechanics.
\medskip

In conclusion, we have found new effects 
in the small effective coupling expansion 
of the one-loop energy shift \eqref{eq:NewTerms}; 
these might be interpreted
as a resolution of the three-loop puzzle. 
We have also derived parts of the first quantum correction 
to the string scattering phase. This is given by the 
interpolating function \eqref{eq:C2} for the dressing phase $\theta$
within the string Bethe ansatz. 
It would be important to understand how this phase behaves for 
finite values of $\lambda$, not just for small or strong coupling. 
In view of many exact results for scattering phases in sigma models,
e.g.,~\cite{Zamolodchikov:1978xm,Wiegmann:1984ec,Faddeev:1985qu}
(see also \cite{Mann:2005ab} in the present context), this is not a hopeless goal. 
Also,  the above argument of self-consistent quantum corrections 
seems suggestive in this direction.

\paragraph{\footnotesize Acknowledgements.}

\footnotesize
We would like to thank D.~Berenstein, S.~Frolov, J.~Maldacena, A.~Mikhailov, J.~Polchinski,
M.~Staudacher and K.~Zarembo for useful discussions of related issues. 
A.A.T.~is grateful to S.\ Sch\"afer-Nameki, M.~Zamaklar and K.~Zarembo 
for discussions on the structure 
of the sums in the one-loop string correction.
The work of N.B.~is supported in part by the U.S.~National Science
Foundation Grant No.~PHY02-43680. 
The work of A.A.T.~was supported by the DOE
grant DE-FG02-91ER40690 and also by the INTAS contract 03-51-6346
and the RS Wolfson award.
Most of this work was carried out while we were participants of the 
KITP program ``Mathematical Structures in String Theory{}''. 
We thank KITP for hospitality  
and acknowledge partial support of NSF grant PHY99-07949 while at KITP. 
Any opinions, findings and conclusions or recommendations expressed in this
material are those of the authors and do not necessarily reflect the
views of the National Science Foundation.

\normalsize

\newpage 
\section*{Appendix}

Here we present some lengthy expressions which arise 
in the sum over frequencies $e(n)$.
The exact expression for $e(n)$ is given in 
\cite{Park:2005ji,Beisert:2005mq,Schafer-Nameki:2005tn}
\<\label{eq:eofn}
e(n)\eq
\frac{1}{4\kappa}\lrbrk{\omega_1+\omega_2-\omega_3-\omega_4}
+\frac{1}{\kappa}\sqrt{n^2+\kappa^2}
+\frac{2}{\kappa}\sqrt{n^2+\cJ^2-m^2}
\nl
-\frac{2}{\kappa}\sqrt{(n-\gamma)^2+\half(\kappa^2+\cJ^2-m^2)}
-\frac{2}{\kappa}\sqrt{(n+\gamma)^2+\half(\kappa^2+\cJ^2-m^2)}\,.
\>
Here the first two terms correspond to bosonic fluctuations along $AdS_5$,
the third to bosonic fluctuations along $S^5$ and the remaining two
to fermionic fluctuations. The frequencies $\omega_1,\ldots,\omega_4$ are
the solutions to the quartic equation
\[
(\omega^2-n^2)^2-\frac{4\cJ m \kappa^2\omega^2}{k\sqrt{\kappa^2+k^2}}
-4\lrbrk{1-\frac{\cJ m}{k\sqrt{\kappa^2+k^2}}}\bigbrk{\omega\sqrt{\kappa^2+k^2}-k n}^2=0
\]
ordered in magnitude from largest to smallest. The shift $\gamma$ is given by
\[
\gamma=\frac{\kappa m}{\sqrt{\kappa^2+k^2}}\,
\frac{\kappa^2-\cJ^2+k^2}{\kappa^2-\cJ^2+m^2}
\]
and, finally, $\kappa$ is determined by the equation%
%
\[\bigbrk{\kappa^2-\cJ^2-m^2}\sqrt{\kappa^2+k^2}+2\cJ km=0.\]

When we expand for large $\cJ$ 
assuming $n=\cJ x$ to be of the order $\cJ$, we obtain 
%
\<\label{eq:IntFull}
e\supup{int}(x)\eq
\frac{(k-m)^2}{32\cJ^4x^2(1+x^2)^{7/2}}\,
\Bigl[
-16m^2+(-3k^2+14km-75m^2)\,x^2
\nl\qquad\qquad\qquad\qquad
+(12k^2-32km+60m^2)\,x^4+(-16km-16m^2)\,x^6
\Bigr]
\nl
+\frac{(k-m)^2}{256\cJ^6x^4(1+x^2)^{11/2}}\,
\Bigl[
(-256km^3+256m^4)
\nl\qquad\qquad\qquad
+(64k^2m^2-1536km^3+1344m^4)\,x^2
\nl\qquad\qquad\qquad
+(15k^4-248k^3m+1118k^2m^2-4624km^3+2907m^4)\,x^4
\nl\qquad\qquad\qquad
+(-180k^4+1420k^3m-2168k^2m^2-3204km^3+2276m^4)\,x^6
\nl\qquad\qquad\qquad
+(120k^4+568k^3m-1892k^2m^2-1728km^3+1076m^4)\,x^8
\nl\qquad\qquad\qquad
+(224k^3m-688k^2m^2-736km^3+368m^4)\,x^{10}
\nl\qquad\qquad\qquad
+(64k^3m-128k^2m^2-128km^3+64m^4)\,x^{12}
\Bigr]+\order{1/\cJ^8}.
\>
Its singular part is given by
\[\label{eq:IntSing}
e\supup{int}\indup{sing}(x)=
-\frac{(k-m)^2 m^2}{2\cJ^4x^2}
-\frac{(k-m)^3 m^3}{\cJ^6x^4}
+\frac{(k-m)^2m^2}{4\cJ^6x^2}\,(k^2-2km-m^2)
+\order{1/\cJ^8}.
\]
We also state the large-$n$ asymptotics of $e\supup{int}(x)$
which is agreement with $e(\cJ x)$
\<\label{eq:IntAsymp}
e\supup{int}(x)=
-\frac{(k-m)^2(k+m)m}{x^3}
\lrbrk{-\frac{1}{2\cJ^4}+
\frac{k^2-3km+m^2}{4\cJ^6}+\order{1/\cJ^8}
}
+\order{1/x^4}.
\>
%

Now we assume $n$ to be fixed and expand.
The regular part in this case was
found in \cite{Schafer-Nameki:2005tn}
\<\label{eq:SumReg}
e\supup{sum}\indup{reg}(n)
\eq
\frac{\bigbrk{n^4-4(k-m)mn^2}^{1/2}}{4\cJ^2}
-
\frac{1}{4\cJ^2}\Bigl[n^2+\bigl(-2km+2m^2\bigr)\Bigr]
\nl
-\frac{\bigbrk{n^4-4(k-m)mn^2}^{-1/2}}{16\cJ^4}
\Bigl[
n^6
+\bigl(6k^2-22km+12m^2\bigr)\,n^4
\nl\qquad\qquad\qquad\qquad\qquad\qquad\qquad
+\bigl(-20k^3m+80k^2m^2-84km^3+24m^4\bigr)\,n^2
\Bigr]
\nl\qquad
+\frac{1}{16\cJ^4}
\Bigl[n^4
+\bigl(6k^2-20km+10m^2\bigr)\,n^2 
\nl\qquad\qquad\qquad\qquad
+\bigl(-8k^3m+30k^2m^2-28km^3+6m^4\bigr)
\Bigr]
\nl
+\frac{\bigbrk{n^4-4(k-m)mn^2}^{-3/2}}{32\cJ^6}
\nl\qquad
\times\Bigl[n^{12}+
\bigl(15k^2-44km+25m^2\bigr)\,n^{10}
\nl\qquad\qquad\qquad
+\bigl(15k^4-218k^3m+603k^2m^2-556km^3+164m^4\bigr)\, n^8 
\nl\qquad\qquad\qquad
+\bigl(-106k^5m+1068k^4m^2-3128k^3m^3
\nl\qquad\qquad\qquad\qquad\qquad
   +3796k^2m^4-2014km^5+384m^6\bigr)\,n^6 
\nl\qquad\qquad\qquad
+\bigl(180k^6m^2-1656k^5m^3+5256k^4m^4-7744k^3m^5
\nl\qquad\qquad\qquad\qquad\qquad
   +5684k^2m^6-1960km^7+240m^8\bigr)\,n^4 
\Bigr]
\nl\qquad
-\frac{1}{32\cJ^6}
\Bigl[
n^6
+\bigl(15k^2-38km+19m^2\bigr)\,n^4 
\nl\qquad\qquad\qquad\qquad
+\bigl(15k^4-128k^3m+279k^2m^2-202km^3+44m^4\bigr)\,n^2 
\nl\qquad\qquad\qquad\qquad
+\bigl(-16k^5m+120k^4m^2-282k^3m^3
\nl\qquad\qquad\qquad\qquad\qquad\qquad
+262k^2m^4-94km^5+10m^6\bigr)
\Bigr]
+\order{1/\cJ^8}
\>
and the singular part reads
\<\label{eq:SumSing}
e\supup{sum}\indup{sing}(n)\eq
-\frac{(k-m)^2}{32\cJ^4}\,(3k^2-16km+19m^2)
+\frac{(k-m)^2n^2}{64\cJ^6}\,(45k^2-162km+153m^2)
\nl
+\frac{(k-m)^2}{256\cJ^6}\,(15k^4-248k^3m+766k^2m^2-752km^3+91m^4)
+\order{1/\cJ^8}.
\>

It can be verified that
$e\supup{int}\indup{sing}(x)=e\supup{sum}\indup{reg}(\cJ x)$
and $e\supup{sum}\indup{sing}(n)=e\supup{int}\indup{reg}(n/\cJ)$,
at least as far as the expansion goes.


\newpage

\bibliographystyle{nbshort}
\bibliography{Jfive}

\begin{thebibliography}{10m}
\ifx\href\asklfhas\newcommand{\href}[2]{#2}\fi
\raggedright
\footnotesize
\parskip 0pt
\parindent -1em
\itemindent -1em
\itemsep 0pt

\bibitem{Gubser:2002tv}
S.~S.~Gubser, I.~R.~Klebanov and A.~M.~Polyakov,
\textsf{Nucl.~Phys.~B636,~99~(2002)},
\href{http://arXiv.org/abs/hep-th/0204051}{\texttt{hep-th/0204051}}.
%
\bibitem{Frolov:2002av}
S.~Frolov and A.~A.~Tseytlin,
\textsf{JHEP~0206,~007~(2002)},
\href{http://arXiv.org/abs/hep-th/0204226}{\texttt{hep-th/0204226}}.
%
\bibitem{Russo:2002sr}
J.~G.~Russo,
\textsf{JHEP~0206,~038~(2002)},
\href{http://arXiv.org/abs/hep-th/0205244}{\texttt{hep-th/0205244}}.
%
\bibitem{Minahan:2002rc}
J.~A.~Minahan,
\textsf{Nucl.~Phys.~B648,~203~(2003)},
\href{http://arXiv.org/abs/hep-th/0209047}{\texttt{hep-th/0209047}}.
%
\bibitem{Frolov:2003qc}
S.~Frolov and A.~A.~Tseytlin,
\textsf{Nucl.~Phys.~B668,~77~(2003)},
\href{http://arXiv.org/abs/hep-th/0304255}{\texttt{hep-th/0304255}}.
%
\bibitem{Frolov:2003xy}
S.~Frolov and A.~A.~Tseytlin,
\textsf{Phys.~Lett.~B570,~96~(2003)},
\href{http://arXiv.org/abs/hep-th/0306143}{\texttt{hep-th/0306143}}.
%
\bibitem{Tseytlin:2003ii}
A.~A.~Tseytlin,
\href{http://arXiv.org/abs/hep-th/0311139}{\texttt{hep-th/0311139}}.
%
\bibitem{Beisert:2003xu}
N.~Beisert, J.~A.~Minahan, M.~Staudacher and K.~Zarembo,
\textsf{JHEP~0309,~010~(2003)},
\href{http://arXiv.org/abs/hep-th/0306139}{\texttt{hep-th/0306139}}.
%
\bibitem{Beisert:2003ea}
N.~Beisert, S.~Frolov, M.~Staudacher and A.~A.~Tseytlin,
\textsf{JHEP~0310,~037~(2003)},
\href{http://arXiv.org/abs/hep-th/0308117}{\texttt{hep-th/0308117}}.
%
\bibitem{Beisert:2004ry}
N.~Beisert,
\textsf{Phys.~Rept.~405,~1~(2005)},
\href{http://arXiv.org/abs/hep-th/0407277}{\texttt{hep-th/0407277}}.
%
\bibitem{Beisert:2004yq}
N.~Beisert,
\textsf{Comptes~Rendus~Physique~5,~1039~(2004)},
\href{http://arXiv.org/abs/hep-th/0409147}{\texttt{hep-th/0409147}}.
%
\bibitem{Zarembo:2004hp}
K.~Zarembo,
\textsf{Comptes~Rendus~Physique~5,~1081~(2004)},
\href{http://arXiv.org/abs/hep-th/0411191}{\texttt{hep-th/0411191}}.
%
\bibitem{Minahan:2002ve}
J.~A.~Minahan and K.~Zarembo,
\textsf{JHEP~0303,~013~(2003)},
\href{http://arXiv.org/abs/hep-th/0212208}{\texttt{hep-th/0212208}}.
%
\bibitem{Beisert:2003tq}
N.~Beisert, C.~Kristjansen and M.~Staudacher,
\textsf{Nucl.~Phys.~B664,~131~(2003)},
\href{http://arXiv.org/abs/hep-th/0303060}{\texttt{hep-th/0303060}}.
%
\bibitem{Beisert:2003yb}
N.~Beisert and M.~Staudacher,
\textsf{Nucl.~Phys.~B670,~439~(2003)},
\href{http://arXiv.org/abs/hep-th/0307042}{\texttt{hep-th/0307042}}.
%
\bibitem{Beisert:2003ys}
N.~Beisert,
\textsf{Nucl.~Phys.~B682,~487~(2004)},
\href{http://arXiv.org/abs/hep-th/0310252}{\texttt{hep-th/0310252}}.
%
\bibitem{Mandal:2002fs}
G.~Mandal, N.~V.~Suryanarayana and S.~R.~Wadia,
\textsf{Phys.~Lett.~B543,~81~(2002)},
\href{http://arXiv.org/abs/hep-th/0206103}{\texttt{hep-th/0206103}}.
%
\bibitem{Bena:2003wd}
I.~Bena, J.~Polchinski and R.~Roiban,
\textsf{Phys.~Rev.~D69,~046002~(2004)},
\href{http://arXiv.org/abs/hep-th/0305116}{\texttt{hep-th/0305116}}.
%
\bibitem{Serban:2004jf}
D.~Serban and M.~Staudacher,
\textsf{JHEP~0406,~001~(2004)},
\href{http://arXiv.org/abs/hep-th/0401057}{\texttt{hep-th/0401057}}.
%
\bibitem{Beisert:2004hm}
N.~Beisert, V.~Dippel and M.~Staudacher,
\textsf{JHEP~0407,~075~(2004)},
\href{http://arXiv.org/abs/hep-th/0405001}{\texttt{hep-th/0405001}}.
%
\bibitem{Staudacher:2004tk}
M.~Staudacher,
\textsf{JHEP~0505,~054~(2005)},
\href{http://arXiv.org/abs/hep-th/0412188}{\texttt{hep-th/0412188}}.
%
\bibitem{Beisert:2005fw}
N.~Beisert and M.~Staudacher,
\textsf{Nucl.~Phys.~B727,~1~(2005)},
\href{http://arXiv.org/abs/hep-th/0504190}{\texttt{hep-th/0504190}}.
%
\bibitem{Kazakov:2004qf}
V.~A.~Kazakov, A.~Marshakov, J.~A.~Minahan and K.~Zarembo,
\textsf{JHEP~0405,~024~(2004)},
\href{http://arXiv.org/abs/hep-th/0402207}{\texttt{hep-th/0402207}}.
%
\bibitem{Kazakov:2004nh}
V.~A.~Kazakov and K.~Zarembo,
\textsf{JHEP~0410,~060~(2004)},
\href{http://arXiv.org/abs/hep-th/0410105}{\texttt{hep-th/0410105}}.
%
\bibitem{Beisert:2004ag}
N.~Beisert, V.~A.~Kazakov and K.~Sakai,
\href{http://arXiv.org/abs/hep-th/0410253}{\texttt{hep-th/0410253}}.
%
\bibitem{Beisert:2005bm}
N.~Beisert, V.~Kazakov, K.~Sakai and K.~Zarembo,
\href{http://arXiv.org/abs/hep-th/0502226}{\texttt{hep-th/0502226}}.
%
\bibitem{Arutyunov:2004vx}
G.~Arutyunov, S.~Frolov and M.~Staudacher,
\textsf{JHEP~0410,~016~(2004)},
\href{http://arXiv.org/abs/hep-th/0406256}{\texttt{hep-th/0406256}}.
%
\bibitem{Schafer-Nameki:2004ik}
S.~Sch{\"a}fer-Nameki,
\textsf{Nucl.~Phys.~B714,~3~(2005)},
\href{http://arXiv.org/abs/hep-th/0412254}{\texttt{hep-th/0412254}}.
%
\bibitem{Beisert:2005di}
N.~Beisert, V.~A.~Kazakov, K.~Sakai and K.~Zarembo,
\textsf{JHEP~0507,~030~(2005)},
\href{http://arXiv.org/abs/hep-th/0503200}{\texttt{hep-th/0503200}}.
%
\bibitem{Callan:2003xr}
C.~G.~Callan,~Jr., H.~K.~Lee, T.~McLoughlin, J.~H.~Schwarz, I.~Swanson and
  X.~Wu,
\textsf{Nucl.~Phys.~B673,~3~(2003)},
\href{http://arXiv.org/abs/hep-th/0307032}{\texttt{hep-th/0307032}}.
%
\bibitem{Tseytlin:2004xa}
A.~A.~Tseytlin,
\href{http://arXiv.org/abs/hep-th/0409296}{\texttt{hep-th/0409296}}.
%
\bibitem{Plefka:2005bk}
J.~Plefka,
\href{http://arXiv.org/abs/hep-th/0507136}{\texttt{hep-th/0507136}}.
%
\bibitem{Maldacena:1998re}
J.~M.~Maldacena,
\textsf{Adv.~Theor.~Math.~Phys.~2,~231~(1998)},
\href{http://arXiv.org/abs/hep-th/9711200}{\texttt{hep-th/9711200}}.
%
\bibitem{Gubser:1998bc}
S.~S.~Gubser, I.~R.~Klebanov and A.~M.~Polyakov,
\textsf{Phys.~Lett.~B428,~105~(1998)},
\href{http://arXiv.org/abs/hep-th/9802109}{\texttt{hep-th/9802109}}.
%
\bibitem{Witten:1998qj}
E.~Witten,
\textsf{Adv.~Theor.~Math.~Phys.~2,~253~(1998)},
\href{http://arXiv.org/abs/hep-th/9802150}{\texttt{hep-th/9802150}}.
%
\bibitem{Beisert:2005mq}
N.~Beisert, A.~A.~Tseytlin and K.~Zarembo,
\textsf{Nucl.~Phys.~B715,~190~(2005)},
\href{http://arXiv.org/abs/hep-th/0502173}{\texttt{hep-th/0502173}}.
%
\bibitem{Hernandez:2005nf}
R.~Hern\'andez, E.~L\'opez, A.~Peri\'a\~nez and G.~Sierra,
\textsf{JHEP~0506,~011~(2005)},
\href{http://arXiv.org/abs/hep-th/0502188}{\texttt{hep-th/0502188}}.
%
\bibitem{Beisert:2005bv}
N.~Beisert and L.~Freyhult,
\textsf{Phys.~Lett.~B622,~343~(2005)},
\href{http://arXiv.org/abs/hep-th/0506243}{\texttt{hep-th/0506243}}.
%
\bibitem{Park:2005ji}
I.~Y.~Park, A.~Tirziu and A.~A.~Tseytlin,
\textsf{JHEP~0503,~013~(2005)},
\href{http://arXiv.org/abs/hep-th/0501203}{\texttt{hep-th/0501203}}.
%
\bibitem{Schafer-Nameki:2005tn}
S.~Sch{\"a}fer-Nameki, M.~Zamaklar and K.~Zarembo,
\href{http://arXiv.org/abs/hep-th/0507189}{\texttt{hep-th/0507189}}.
%
\bibitem{Frolov:2004bh}
S.~A.~Frolov, I.~Y.~Park and A.~A.~Tseytlin,
\textsf{Phys.~Rev.~D71,~026006~(2005)},
\href{http://arXiv.org/abs/hep-th/0408187}{\texttt{hep-th/0408187}}.
%
\bibitem{Schafer-Nameki:2005is}
S.~Schafer-Nameki and M.~Zamaklar,
\href{http://arXiv.org/abs/hep-th/0509096}{\texttt{hep-th/0509096}}.
%
\bibitem{Arutyunov:2003za}
G.~Arutyunov, J.~Russo and A.~A.~Tseytlin,
\textsf{Phys.~Rev.~D69,~086009~(2004)},
\href{http://arXiv.org/abs/hep-th/0311004}{\texttt{hep-th/0311004}}.
%
\bibitem{Frolov:2003tu}
S.~Frolov and A.~A.~Tseytlin,
\textsf{JHEP~0307,~016~(2003)},
\href{http://arXiv.org/abs/hep-th/0306130}{\texttt{hep-th/0306130}}.
%
\bibitem{Klebanov:2002mp}
I.~R.~Klebanov, M.~Spradlin and A.~Volovich,
\textsf{Phys.~Lett.~B548,~111~(2002)},
\href{http://arXiv.org/abs/hep-th/0206221}{\texttt{hep-th/0206221}}.
%
\bibitem{Mann:2005ab}
N.~Mann and J.~Polchinski,
\href{http://arXiv.org/abs/hep-th/0508232}{\texttt{hep-th/0508232}}.
%
\bibitem{Klose:2005aa}
T.~Klose,
PhD thesis.
%
\bibitem{Beisert:2005aa}
N.~Beisert and T.~Klose,
work in progress.
%
\bibitem{Arutyunov:2003rg}
G.~Arutyunov and M.~Staudacher,
\textsf{JHEP~0403,~004~(2004)},
\href{http://arXiv.org/abs/hep-th/0310182}{\texttt{hep-th/0310182}}.
%
\bibitem{Inozemtsev:2002vb}
V.~I.~Inozemtsev,
\textsf{Phys.~Part.~Nucl.~34,~166~(2003)},
\href{http://arXiv.org/abs/hep-th/0201001}{\texttt{hep-th/0201001}}.
%
\bibitem{Berenstein:2002jq}
D.~Berenstein, J.~M.~Maldacena and H.~Nastase,
\textsf{JHEP~0204,~013~(2002)},
\href{http://arXiv.org/abs/hep-th/0202021}{\texttt{hep-th/0202021}}.
%
\bibitem{Eden:2004ua}
B.~Eden, C.~Jarczak and E.~Sokatchev,
\textsf{Nucl.~Phys.~B712,~157~(2005)},
\href{http://arXiv.org/abs/hep-th/0409009}{\texttt{hep-th/0409009}}.
%
\bibitem{Fischbacher:2004iu}
T.~Fischbacher, T.~Klose and J.~Plefka,
\textsf{JHEP~0502,~039~(2005)},
\href{http://arXiv.org/abs/hep-th/0412331}{\texttt{hep-th/0412331}}.
%
\bibitem{Minahan:2004ds}
J.~A.~Minahan,
\textsf{JHEP~0410,~053~(2004)},
\href{http://arXiv.org/abs/hep-th/0405243}{\texttt{hep-th/0405243}}.
%
\bibitem{Engquist:2003rn}
J.~Engquist, J.~A.~Minahan and K.~Zarembo,
\textsf{JHEP~0311,~063~(2003)},
\href{http://arXiv.org/abs/hep-th/0310188}{\texttt{hep-th/0310188}}.
%
\bibitem{Freyhult:2005fn}
L.~Freyhult and C.~Kristjansen,
\textsf{JHEP~0505,~043~(2005)},
\href{http://arXiv.org/abs/hep-th/0502122}{\texttt{hep-th/0502122}}.
%
\bibitem{Kristjansen:2004ei}
C.~Kristjansen,
\textsf{Phys.~Lett.~B586,~106~(2004)},
\href{http://arXiv.org/abs/hep-th/0402033}{\texttt{hep-th/0402033}}.
%
\bibitem{Kristjansen:2004za}
C.~Kristjansen and T.~M\r{a}nsson,
\textsf{Phys.~Lett.~B596,~265~(2004)},
\href{http://arXiv.org/abs/hep-th/0406176}{\texttt{hep-th/0406176}}.
%
\bibitem{Swanson:2004mk}
I.~Swanson,
\href{http://arXiv.org/abs/hep-th/0405172}{\texttt{hep-th/0405172}}.
%
\bibitem{Kruczenski:2003gt}
M.~Kruczenski,
\textsf{Phys.~Rev.~Lett.~93,~161602~(2004)},
\href{http://arXiv.org/abs/hep-th/0311203}{\texttt{hep-th/0311203}}.
%
\bibitem{Kruczenski:2004kw}
M.~Kruczenski, A.~V.~Ryzhov and A.~A.~Tseytlin,
\textsf{Nucl.~Phys.~B692,~3~(2004)},
\href{http://arXiv.org/abs/hep-th/0403120}{\texttt{hep-th/0403120}}.
%
\bibitem{Mikhailov:2004xw}
A.~Mikhailov,
\textsf{JHEP~0409,~068~(2004)},
\href{http://arXiv.org/abs/hep-th/0404173}{\texttt{hep-th/0404173}}.
%
\bibitem{Zamolodchikov:1978xm}
A.~B.~Zamolodchikov and A.~B.~Zamolodchikov,
\textsf{Annals~Phys.~120,~253~(1979)}.
%
\bibitem{Wiegmann:1984ec}
P.~Wiegmann,
\textsf{Phys.~Lett.~B142,~173~(1984)}.
%
\bibitem{Faddeev:1985qu}
L.~D.~Faddeev and N.~Y.~Reshetikhin,
\textsf{Ann.~Phys.~167,~227~(1986)}.
%
\end{thebibliography}

\end{document}